\newcommand{\be}{\begin{equation}}
\newcommand{\ee}{\end{equation}}
\newcommand{\bea}{\begin{eqnarray}}
\newcommand{\eea}{\end{eqnarray}}

\newcommand{\nn}{\nonumber}

\def\p{\partial }

\def\a{\alpha }

\def\s{\sigma }

\def\ov{\over  }

\def\ga{\alpha }

\def\I{{\rm I}}
\def\II{{\rm II}}

\documentclass[12pt]{article}

\input epsf
\usepackage{amsmath,amsfonts,epsfig,color,latexsym}
\usepackage{amssymb}
\usepackage{amsfonts}
\usepackage{amssymb}
\renewcommand{\thefootnote}{\fnsymbol{footnote}}

\def\appendix#1{
  \addtocounter{section}{1}
  \setcounter{equation}{0}
  \renewcommand{\thesection}{\Alph{section}}
  \section*{Appendix \thesection\protect\indent \parbox[t]{11.15cm}
  {#1} }
  \addcontentsline{toc}{section}{Appendix \thesection\ \ \ #1}
  }

\begin{document}


\null\vskip-24pt 
\hfill {\tt hep-th/0606110}
\vskip0.2truecm
\begin{center}
\vskip 0.2truecm {\Large\bf 
Black hole formation from collisions \\
of cosmic fundamental strings}
\vskip 0.2truecm

\vskip 0.7truecm
\vskip 0.7truecm

{\bf Roberto Iengo$^a$ and Jorge G. Russo$^b$}\\
\vskip 0.4truecm
\vskip 0.4truecm

${}^a${\it  International School for Advanced Studies (SISSA)\\
Via Beirut 2-4, I-34013 Trieste, Italy} \\
{\it  INFN, Sezione di Trieste}

\medskip

$^{b}${\it 
Instituci\' o Catalana de Recerca i Estudis Avan\c{c}ats (ICREA),\\
Departament ECM,
Facultat de F\'\i sica, Universitat de Barcelona,
 Spain} 
\end{center}
\vskip 0.2truecm 

\noindent\centerline{\bf Abstract}

We develop the general formalism for joining, splitting and
interconnection of closed and open strings.
As an application, we study  examples of  fundamental cosmic string collisions 
leading to gravitational collapse. We find that the interconnection of two
strings of equal and opposite maximal angular momentum  
and arbitrarily large mass 
generically leads to the formation of black holes, 
provided their relative velocity is small 
enough. 

\newpage

\renewcommand{\thefootnote}{\arabic{footnote}}
\setcounter{footnote}{0}

\section{Introduction}

Recent works have indicated that in brane inflation models
cosmic strings are copiously produced during the brane collision
\cite{tye1,tye2}.
This has led to a renewed interest in the physics of cosmic strings
and to consider the exciting possibility that there could be long-lived
fundamental strings of cosmic size 
(for reviews, see \cite{kibble1,polchinski,tye3,kibble2} 
and references therein). 
Finding such objects  could constitute a test of string theory.

The dynamics of cosmic strings could lead to interesting astrophysical events
such as gravitational waves or black hole formation.
Some aspects of the dynamics of cosmic string interactions were studied in
\cite{myers,copeland} (for Abrikosov-Nielsen-Olesen strings, see
recently  
\cite{achucarro}).

Here we will develop in full detail 
the classical formalism of string splitting, joining
and intercommutation. Our formulas (Appendix)
provide explicit expressions for 
the outgoing string solution starting with an arbitrary initial string
configuration before interaction.
Understanding the dynamics and the different features of 
splitting and joining processes
is of interest, since these processes are the basis of the interaction
rules in string theory.

As an application, we will study the process of possible gravitational
collapse as a result of the collision of cosmic fundamental strings.
Surprisingly, we will find that gravitational collapse is a quite
common phenomenon ensuing  the encounter of strings of equal and opposite 
maximal angular momentum, which classically are rotating straight strings,
and folded in the case of the closed string.
If the initial strings just touch at the end points, then they can join 
forming one single string. If they meet at some intermediate point,
then 
they can interconnect giving rise to two new strings.   

We then study the evolution of the resulting strings by the standard 
flat spacetime dynamics. We find that they typically contract in a finite time to a
minimum size, which sometimes is smaller than the gravitational radius $R_s$.

If the strings meet with zero relative transverse momentum, we find that, 
for generic values of the intersecting positions and angle, a finite
fraction of the mass of the resulting interconnected strings collapses
into a mathematical point.  

In any of these situations, gravitational forces become very strong
when the string size approach $R_s$ and should enhance the evolution
towards the collapse, ensuing in the formation of a horizon and hence
a black hole \cite{hawking,polmarev,vilenkin} (other  discussions can be
found e.g. in \cite{DV,matsuda}).
In our computation the mass (proportional to the length) of the strings appear as an overall scale, therefore this phenomenon can occur for arbitrarily large values of the mass. 

If the transverse momentum is not zero, then the resulting strings
will be stretched in the transverse direction. The size is 
of the order of the length times the relative transverse velocity $v$. 
For non-relativistic relative motions with $v$ much less 
than the product of the gravitational constant times the string tension,
this size will be much smaller than the gravitational radius. We conclude that
also in this case, the  interconnection of our strings generically leads to the formation of a black hole.

We also consider a long-lived version of the string with maximum
angular momentum. This is a closed string with some component in extra
dimensions,
whose motion in $3+1$ dimensions is the same as the open (or folded) 
string with maximum angular momentum.
We discuss for which  range of the parameters and for which magnitudes
of cross sections  black hole formation is to be expected.


\section{New examples of long-lived cosmic strings}

We  consider type II strings in the presence of D branes or with
extra dimensions compactified on a torus.
We are interested in constructing cosmological strings
where breaking processes are suppressed, leading to  long-lived configurations.
Such strings decay primarily by emission of soft gravitational
radiation.\footnote{The most stable massive non-BPS closed string
in type II string theories seems to be a circular string rotating in
two or more planes \cite{CIR2}, 
for which breaking is maximally suppressed and radiation
is feeble. We will not study this string in this paper.}

\subsection{Rotating straight string on $M^4\times S^1$}

Let $t,X,Y,Z$ represent uncompact coordinates of $M^4$ (3+1 dimensional Minkowski space) and
$W$ compact dimensions of radius $R$.
The solution is as follows:
\bea
X &=& L\cos\tau\cos\s \ ,\qquad
X_R(\s_- ) = {1\over 2}\, L \cos\s_-\ ,\qquad X_L(\s_+ )={1\over 2}\, L \cos\s_+\ ,
\nn\\
Y &=& L\sin\tau\cos\s \ ,\qquad Y_R(\s_- ) = -{1\over 2}\, L \sin\s_-\ ,\qquad Y_L(\s_+ )={1\over 2}\, L \sin\s_+\ ,
\nn\\
W &=& nR\s \ ,\qquad W_R(\s_- ) = {1\over 2}\, nR  \s_-   \ ,\qquad          W_L(\s_+ )={1\over 2}\, nR\s_+  \ ,
\nn\\
t &=& \kappa \, \tau\ ,\qquad \kappa=\sqrt{L^2 +n^2R^2}\ ,
\eea
where $\s_\pm=\s\pm\tau $, $\s\in [0,2\pi)$ and $n$ is an 
integer representing the winding number.

The solution is classically unbreakable for $n=1$. This can be seen as follows.
The closed string can break only if at some given time (say $\tau=0$) there are two different points of the string which get in contact, i.e. $\vec X(\s_1)= \vec X(\s_2)$ for the non-compact coordinates, and for the compact coordinate $W(\s_1)= W(\s_2)+2\pi k R $ for some $k$.
Using this condition for $X$ and $Y$ we find $\s_1=2\pi - \s_2$.
Now the condition  for $W$ gives $\pi-\s_2=k\pi/n$.
For $n=\pm 1$ this gives $\s_1=\s_2=0$ mod $2\pi$, which means that the string cannot break
because the two points coincide and thus there is no string left between them.
For $|n|>1 $, one always has at least one solution with $\s_1\neq\s_2$, and the string can break.

A well known stable string on $M^ 4\times S^ 1$ is the BPS string 
\cite{och,iglesias} 
$\vec X=\vec X_L(\s_+ )$, $W= nR\s +2\a' m/R$. In our case, the mechanism for
stability is different. Although the string looks folded in $3+1$ dimensions
--~and in fact it looks identical to the unstable rotating string of
maximal angular momentum \cite{CIR,handbook}~--
it cannot break classically because all points of the string are separated in the internal dimension
$W$. If $R\gg \sqrt{\a'}$, then breaking by quantum effects is also suppressed.
As mentioned above, it can decay by radiation, with a rate (in four
dimensions)~\cite{handbook}
$\Gamma \sim g_s^2\,  M$, $M\sim \mu L$, where $\mu=(2\pi\a')^{-1}$ is the
string tension and $g_s$ is the closed string coupling constant. The radiation is dominated by soft modes with emitted
energy  $\omega\sim 1/L$. Thus
\be
-{dM\over dt} \sim  \Gamma \times \omega \cong c_0 g_s^2\mu \ ,
\label{eee}
\ee
where $c_0$ is a numerical constant of order one.
Therefore the string takes a time $\sim M/g_s^2$ (or $\sim L/g_s^2$) to
substantially decrease its mass. 

\subsection{Rotating open string which oscillates in extra dimensions}

Consider a brane-world model, with 
a D3-brane placed in the three uncompact directions $X,Y,Z$ of our universe.
Let $W$ stand for an extra dimension. 
The open string solution with Dirichlet boundary conditions at
$W=0$ is given by
\bea
X &=& L\cos\theta \cos\tau\ \cos\s \ ,\qquad Y=L\sin\tau\ \cos\s\ , 
\nn\\
W &=& L\sin\theta \sin\tau\ \sin\s \ ,\qquad t=L\ \tau\ ,\ \ \
L=2\a' M\ .
\label{squash}
\eea
This solution is the analog of the squashing closed string of \cite{handbook}
in the case of an open string. In fact, it has the same form, but here
$\s\in [0,\pi )$ whereas for the closed string $\s\in [0,2\pi )$.

The solution represents a string rotating in the plane $X,Y$, with the
ends attached on the brane  $W=0$, which at the same time
oscillates in the extra dimension $W$ (with maximum amplitude
$\pm L\sin\theta $ for the middle point).
The string can break only at the special times where it 
lies on the brane $W=0$,
namely $\tau =n\pi $, as otherwise it is impossible to get Neumann conditions 
on the free endpoints in the $W$ direction. The time between two successive 
events, when much of string lies within a width $\sim l_s\equiv\sqrt{\alpha'}$ from the brane,
is of the order of $L$. Therefore, for $L$ large and
assuming $L\sin(\theta)>>l_s$, this string lives a long time $\sim L\sim M$,
before breaking into massive pieces. In the following, we will consider the 
interactions of strings during this time interval, in which the breaking is 
exponentially suppressed, since it is classically forbidden. 

As an aside remark, we can note that during the time $\sim l_s$ 
in which the string lies on the brane, it will break with a probability that using the rules of \cite{handbook}
is found to be $\sim g_o^2L/l_s$, where $g_o$ is the open string coupling constant and $g_o^2=g_s$. 
If this is much less than one, it may take several cycles  (a number of order 
$l_s/(g_o^2L)$~) for the string to break, each cycle lasting a time $\sim L$.

The string nevertheless loses energy at all times by gravitational
radiation.\footnote{Emission of vector bosons can be
shown to be suppressed
with respect to graviton emission by inverse powers of $L$.}
The decay rate by massless emission can be estimated following the analysis of 
\cite{CIR2, handbook}. In these works the classical radiation rate 
was found to be a function of $L$ and of the emitted energy $\omega$ of the
following factorized form:
\be
\qquad Rate(N_0)\sim L^{5-D}\cdot F(N_0)\ .
\ee
where $D$ is the number of uncompact spacetime dimensions and
$F(N_0)$ is a decreasing function of the integer $N_0\equiv L \omega$.
Therefore, for large $L$, the massless emission  is concentrated at 
small $\omega$,
where the classical result is expected to hold.
The highest decay rate occurs in the case of $D=4$, since spaces with 
$D>4$ have a decay rate suppressed  by inverse powers of $L$. 
{}For $D=4$, by summing over $N_0$ we find that
\be
\qquad \Gamma_{\rm graviton}\sim g_s^2 \sqrt{N}\ ,
\qquad N=\a'M^2
\ .\ee
{} Since, as in the previous example of Section 2.1,
the radiation is dominated by soft modes with emitted
energy  $\omega\sim 1/L$,
the lifetime required for a substantial decrease of the energy is
again of order $M/g_s^2$. In spaces with  $D>4$, this string lives
an even  longer time.

\section{Black hole formation}

\subsection{ By joining of strings}

In this subsection we show an example of a process where two long strings
join by their ends and the resulting string becomes very small (in fact, pointlike)
during the evolution.

\smallskip

\setcounter{equation}{0}

First, consider two open strings with maximum angular momentum, zero linear momentum and equal energies described by the solutions:
$X_{\I,\II} (\tau,\s )=X_{\I,\II L}(\tau+\sigma )+ X_{\I,\II L}(\tau-\sigma )$
with
\bea
X_\I (\s,\tau ) &=& L\cos\s\cos\tau \ ,\ \ \ \
Y_\I (\s,\tau ) = L\cos\s\sin\tau \ ,\
\nn\\  
X_{\I L}(s)&=& {L\over 2}\,  \cos s ,\qquad  Y_{\I L}(s) ={L\over 2}\, \sin s
\nn\\
X_\II (\s,\tau )&=& 2L + L\cos\s\cos\tau \ ,\ \ \ \ Y_\II (\s,\tau ) = - L\cos\s\sin\tau \ ,\
\nn\\
X_{\II L}(s)&=& L+ {L\over 2}\,  \cos s ,\qquad  Y_{\I L}(s) =- {L\over 2}\, \sin s
\eea
The strings I and II have equal and opposite angular momenta given by
$J_\I=L^2/\a' $, $J_\II=-L^2/\a' $.
As they rotate, the end $\s=0 $ of the string I touches the end
$\s=\pi $
of the string II at $\tau =n \pi $, $n=$ integer.

Consider the situation where the strings join at $\tau =0$.
The resulting open string solution has $J=0$,
since angular momentum is conserved and the original 
total angular momentum of
the system is zero.
 By applying the formulas of appendix A, we find the
 solution after the joining: $X (\tau,\s )=X_{ L}(\tau+\sigma )+ X_{L}(\tau-\sigma )$
with
\bea
X_L(s) &=& 
\begin{cases}
L+ {L\over 2}\, \cos 2s\  & \ -{\pi\over 2} \leq s< {\pi\over 2} \cr
-{L\over 2}\, \cos 2s\  & \ {\pi\over 2}\leq s< {3\pi\over 2}\cr
\end{cases}
\nn \\
Y_L(s) &=&  -{L\over 2}\, \sin 2s\ .
\label{colla}
\eea 
This solution is shown in figure 1.
It describes an open string which at $\tau=0$ is completely straight, then it bends and
contracts until it becomes a point at $\tau =\pi/2$. The solution is periodic with period $\pi $. 

Note that the joining process can occur at the lowest order in string perturbation theory
(with probability $O(g_o^2)$ once one has got this initial configuration).

\begin{figure}[ht]
\centerline{
\epsfig{file=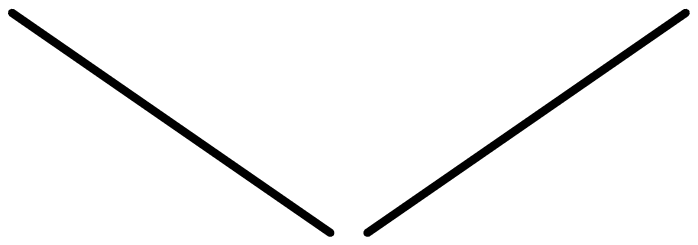,height=.18\textwidth, width=.4\textwidth}\ 
\epsfig{file=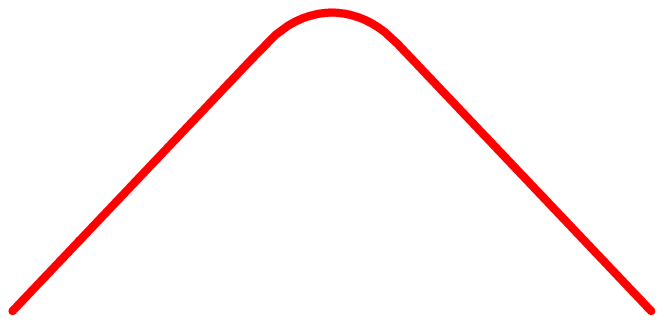,height=.18\textwidth,width=.4\textwidth}\  
\epsfig{file=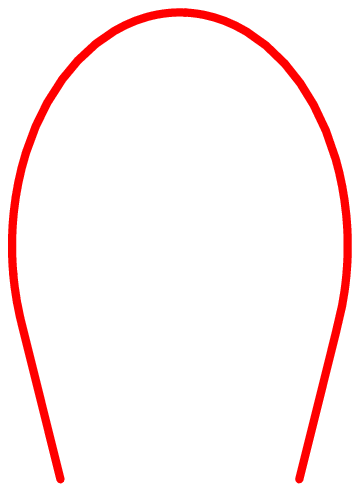,height=.18\textwidth,width=.4\textwidth}
 }
\centerline{
\epsfig{file=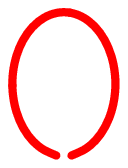,height=.18\textwidth,width=.4\textwidth}\  
\epsfig{file=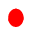,height=.18\textwidth,width=.4\textwidth}\  
\epsfig{file=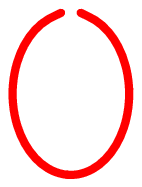,height=.18\textwidth,width=.4\textwidth}
 }

 \caption{\it Evolution of the open string which results from the
 joining
of two open strings with maximum and opposite angular momentum. }
\end{figure}

\bigskip

So far gravitational effects have not been taken into account.
We start with two long straight strings $J_{\rm max} + {\rm anti}J_{\rm max}$ which join forming an open string with $J=0$ in
a regime where its size is much larger than the gravitational radius $R_s$. 
In this regime, the evolution of the string is governed by the classical string equations of motion
in flat spacetime.
As the open string reduces its size, gravitational effects become more and more important. 
A string which reduces to a point
should clearly undergo gravitational collapse. This should happen
when the size of the string becomes smaller than $R_s$. 
For a string of length $\ell \sim M/\mu $, where $\mu =1/(2\pi\a')$ is the string tension,
the gravitational radius is $R_s \sim 2 G\mu \ell $.
 Therefore, when the string contracts by a factor
of order $(G\mu )^{-1}$,  gravitational collapse should be inevitable and a horizon will form \cite{hawking,polmarev,vilenkin}.

One important question is whether the open string could radiate out most of its  energy before its size
becomes smaller than the Schwarzschild radius.
There are two decay channels: the radiation channel, where the string emits a graviton,
and the massive channel, where the string breaks into two pieces.

Let us first consider  the massive channel.
The evolution of the two pieces after the breaking can be followed
by using the general formulas of appendix A. Because of momentum conservation,
in the present case each piece will carry a momentum in the inward
direction. Therefore the system cannot lose energy by breaking. 
This argument ignores gravitational effects.
Taking into account the attractive nature of  gravitational forces 
reinforces the fact that each piece will follow
an inward collapse.

The radiated energy can be estimated with the rules of \cite{handbook}.
For a smooth string in four dimensions (and even if there are kinks), 
the mass loss rate is given by eq. (\ref{eee}).
For this string, $t=2\alpha' M\tau = {M\over \pi\mu }\, \tau $.
As the string loses mass, the value of $M$ is changing. This process is very slow
for large $M$, so we can follow it adiabatically and assume that at each time
the string is described by the same solution with $M(t)$.
We then write $dt={1\over \pi\mu }\, (Md\tau +\tau dM)$.
Hence
\be
-{dM\over  M}  \cong {c_0 g_s^2 d\tau\over \pi +c_0g_s^2 \tau }\ .
\label{eeee}
\ee
The total energy radiated from the initial configuration 
until the string becomes a point is obtained
by integrating this equation from $\tau =0 $ to $\tau =\pi/2 $. We get
\be
M(\tau = {\pi\over 2}) = \beta\, M (\tau = 0)\ ,\qquad
\beta = ( 1+{c_0 g_s^2\over 2} )^{-1}\ . 
\ee
Since $c_0$ is a number of order 1, this mass is of the same order of the initial mass.
This shows that a black hole will be formed before the string becomes a point.

\medskip

The starting point of the above example involves two open strings,
which we know to be unstable.
The same process can occur for the long-lived strings
of Section 2. 

In the case of the joining of two rotating open strings which
oscillate in extra dimensions (Section 2.2),
when the amplitude of oscillation is much smaller than the length of the string
(corresponding to the parameter  $\theta\ll 1 $ in eq.~(\ref{squash})), the dynamics of the joining process is essentially
the same as in the above example (at the same time,  the amplitude of
oscillation must be much larger than $\sqrt{\a'} $ to suppress
breaking by quantum effects).

In the other case, that is considering the joining of two  rotating
straight closed strings on $M^4\times S^1$ with winding number $n=1$
(Section 2.1),
one can easily see that the joining equations for the $X,Y$
coordinates are the same as for the open string case seen 
above. It is also easy to see that the joining conditions are
satisfied
 in the $W$ coordinate,
giving rise to two possible solutions,  with winding 0 or with winding
equal to 2.  

\medskip

As an aside remark, we note that, as far as the flat space evolution is concerned, 
 as the string (\ref{colla}) is contracting, the ends of the string approach each
other,
and they touch only in the limit that the string is a point.
Although in this limit quantum and gravitational effects are important, it is
nevertheless interesting to follow the classical world-sheet evolution. 
When the ends of the string touch,  there is a certain probability
given by the coupling constant $g_o^2$ 
that they  join forming a closed string.
The resulting solution is obtained by
defining $\vec X_L(s)$ and $\vec X_R(s)$ to be equal to the open
string $\vec X_{L,R}$ 
at $\tau=\pi/2$ 
and, for $\tau>\pi/2$, one  imposes closed string boundary
conditions, $\vec X(\tau, \s+2\pi )= \vec X(\tau,\s )$.
Remarkably, the resulting solution is the pulsating circular string solution
described by (we set the center of mass coordinate to zero)
\bea
X(\s ,\tau  ) &=& 2L\, \cos\s\cos\tau\ ,\ \ \ \ Y(\s ,\tau ) = 2L\, \sin\s\cos\tau\ ,\
\nn\\
X_{L}(s)&=& X_{R}(s)= L \cos s ,\qquad  Y_{L}(s)= Y_{R}(s)=L\sin s\ .
\eea

Another remark is that the circular pulsating string can be obtained
from the quantum scattering of two gravitons.
In ref. \cite{handbook} it was remarked that the quantum amplitude for the process 
$two ~ gravitons \leftrightarrow pulsating ~ string$
is the same as the amplitude for $two ~ gravitons \leftrightarrow Jmax ~ string$. 
The rate of  the last process  was computed in 
ref. \cite{kalki} and checked in ref. \cite{IRold}.
{}From these results, we find the rate for the present process
of two gravitons forming a pulsating circular string: 
$\Gamma \sim g^2_se^{-{1\over 2} \a' M^2(\log(4)-1)}$ where $M$ is the mass of the pulsating string.
Once the circular pulsating string is formed, it should inevitably collapse 
into a black hole (if it is not already inside the horizon, it will shrink with a negligible probability of breaking \cite{handbook}). 
 Therefore, this process provides an example of a first-principle calculation based on string 
perturbation theory of black hole formation. The  cross-section for that particular final state
is exponentially small.
In four spacetime dimensions: $\sigma\sim\a'g^2_s e^{-{1\over 2}\a'
M^2 (\log(4)-1)}$.

\subsection{By interconnection of two strings}

In the previous example, black hole formation requires a special initial configuration such that
the endpoints of the two strings touch during the evolution.
A more generic process is the case of string interconnection.\footnote{
{}For open strings, this process corresponds to the u-channel open string diagram (we thank D.~Amati for a discussion on this point).}

When two fundamental strings cross, there is a probability given by the string coupling that
the strings will interconnect, as in fig. 2.
As shown in the figure, there are two possible ways that the string can interconnect.
This is a common process in 3+1 dimensions, where two infinitely long strings always cross for generic initial data. For finite-size strings, the collision has a cross section
of the order of the square of the length of the string.

\bigskip

\begin{figure}[ht]
\centerline{
\epsfig{file=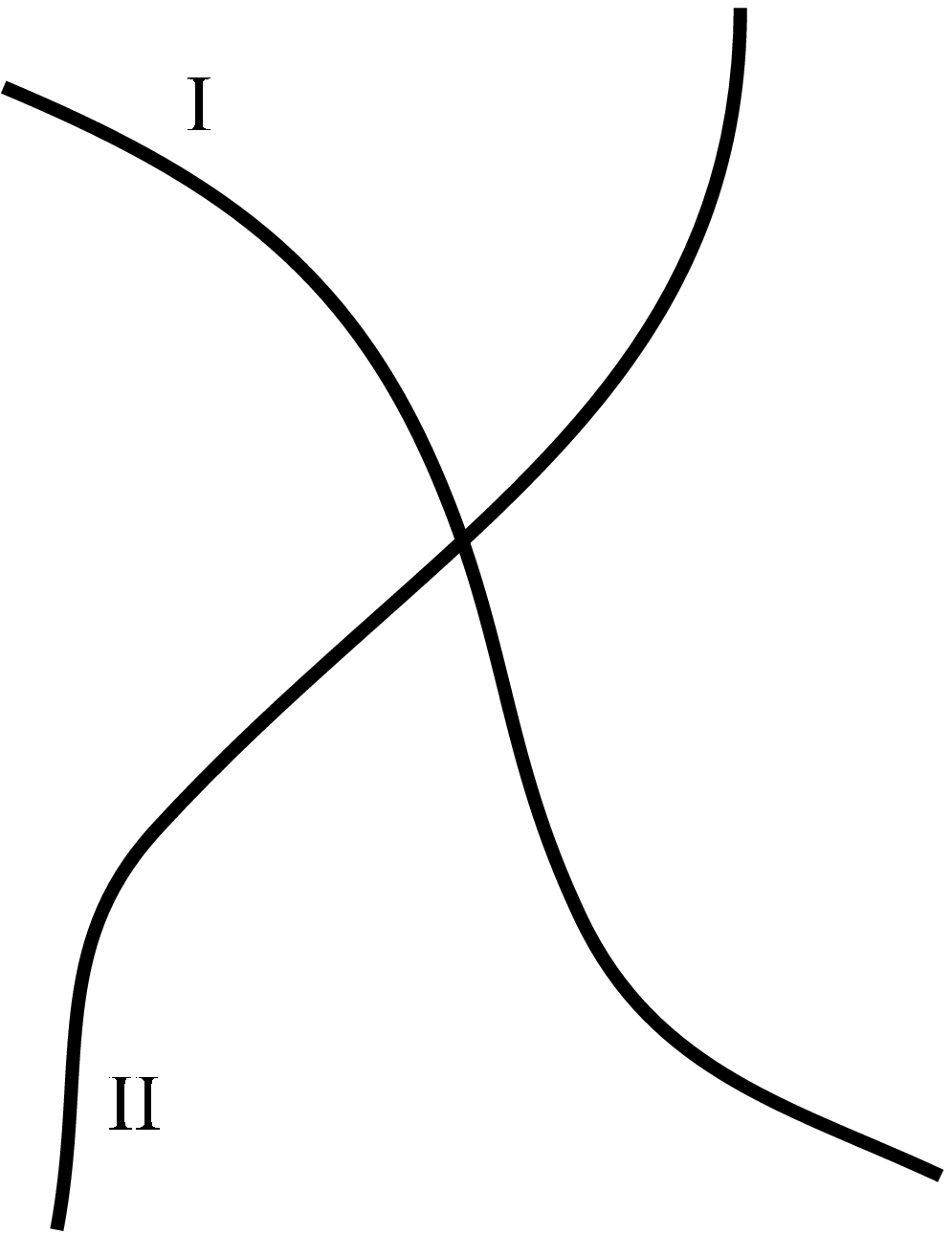,width=.25\textwidth}\qquad \
\ 
\epsfig{file=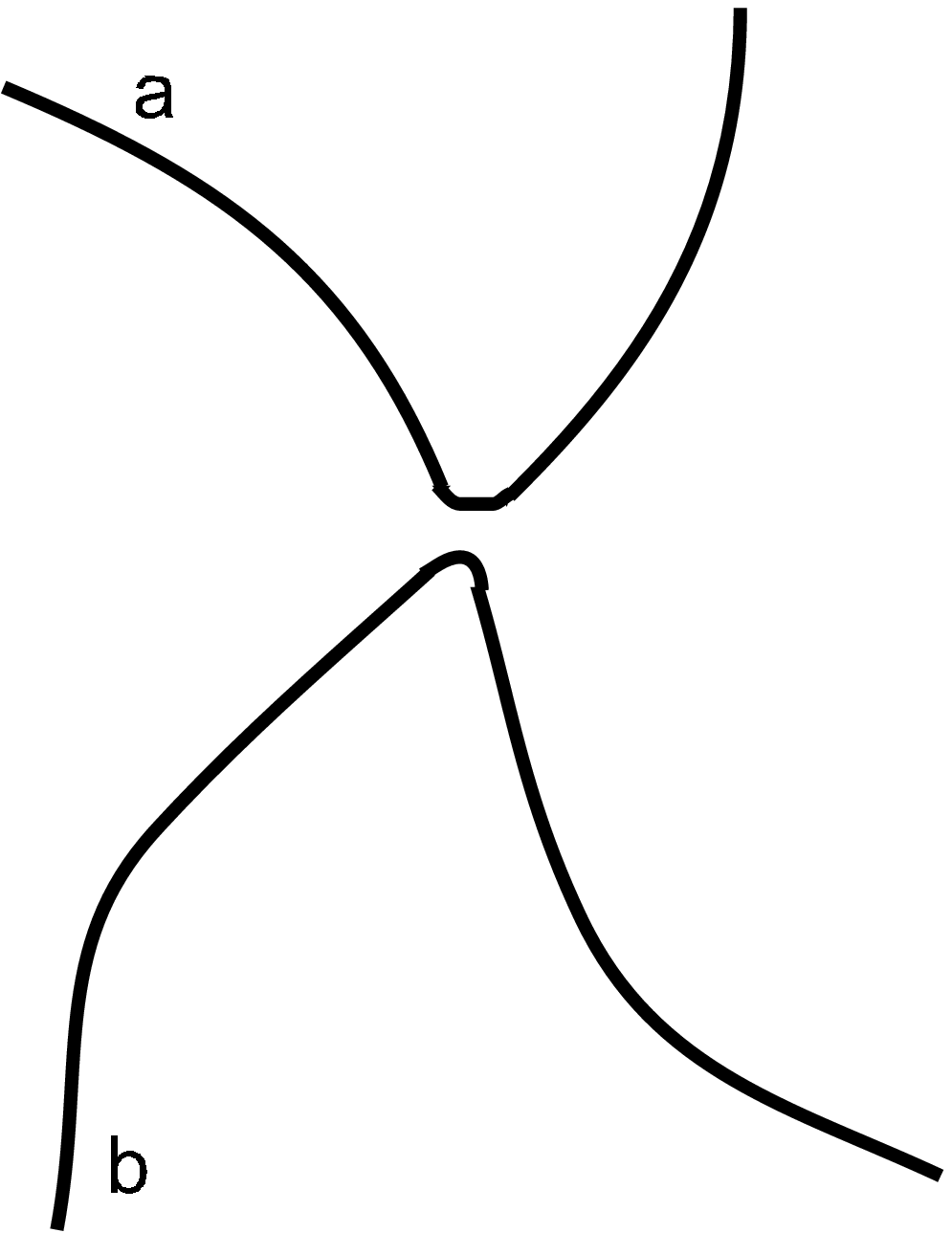,width=.26\textwidth}\ 
\epsfig{file=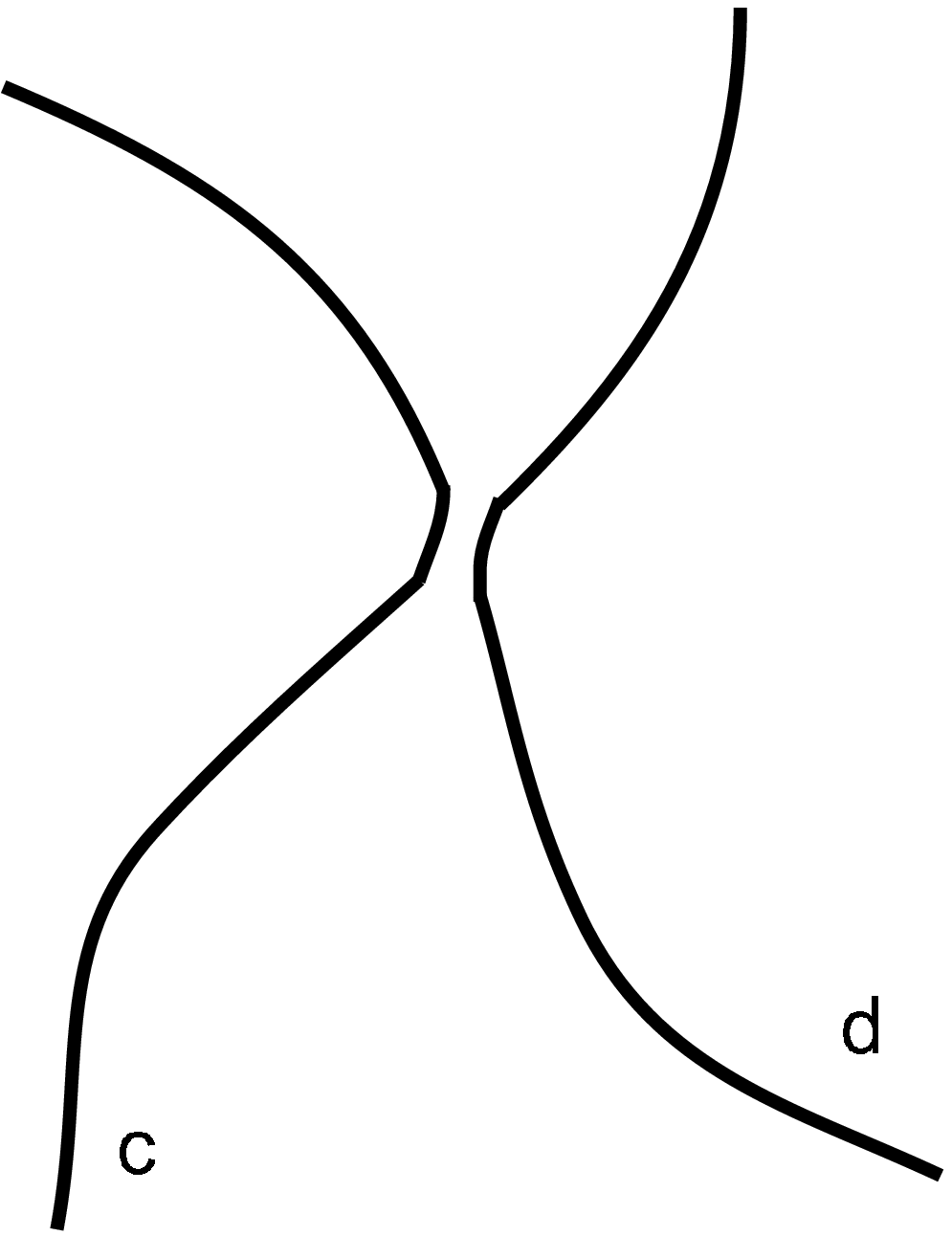,width=.27\textwidth}\ 
}
 \caption{\it Interconnection process. When two strings cross, there
 are two possible ways that they can interconnect, leading to strings
 $a$ and $b$ or strings $c$ and $d$. }
\end{figure}

\bigskip

An interesting question is what is the probability that a black hole is formed
as a result of the collision.
Computing this from string perturbation theory is obviously very complicated,
so we will try to address this question by means of the following experiment:
we send two straight rotating strings against each other,
with random position for the center of mass coordinates and random value
for the relative orientation (within the range where the interconnection is possible).
After repeating the experiment $N_e$ times, we ask how many of the resulting
string configurations are black holes. 
We will consider several conditions
for black hole formation.
One condition is that one of the two final strings completely
lie inside its Schwarzschild radius $R_s$
at some time during the evolution (ignoring detailed features due to the angular momentum).
Another condition is that at some
time the average size of the string lies inside its Schwarzschild radius.
Finally, a third condition, is that a segment of the string
lies within the Schwarzschild radius.
 In our study, the reduction to a
small size just follows
by the natural shrinking of the string that results from flat space
evolution, without taking into account gravity.
In any of these three situations, gravitational forces become very strong
when the string size approach $R_s$ and should enhance the evolution
towards the collapse.

Consider first the interconnection of two rotating open strings of the
type described in section 2.2.
They rotate in the plane $X,Y$, oscillate in the extra $W$ dimension,
and they may also have
transverse momentum in the $Z$ direction on the brane.
We will consider the case of opposite angular momentum in the $X,Y$ plane.
After interconnection, the two emerging strings also spread in the $Z$ direction.
When the center-of-mass transverse $Z$-motion of the string is non-relativistic, 
the spread in the $Z$ direction can be neglected as compared to the spread
in the $X,Y$ direction. In addition, as
 discussed in the previous joining case, 
when the amplitude of oscillation in the $W$ direction is much smaller than the length of the string
(corresponding to the parameter  $\theta\ll 1 $), the dynamics of the interconnection process is essentially
the same as that of the interconnection of two open strings of maximum angular momentum.
Therefore, to simplify the discussion, we will first consider the case
of two open strings with equal and opposite maximum angular momenta
lying at $Z=0$ and 
$W=0$.

\subsubsection{The solutions after the interconnection}

Consider two open strings of (opposite) maximal angular momentum 
in the $XY$ plane, 
having the same energy, which cross
at  some angle at $\tau =0$ . We will take the gauge 
$t=L \tau$. 
The solutions are, with $0\leq s_{\pm}={\s\pm\tau\ov L}\leq 2\pi$,
\bea
X_{\I L} &=& {L\over 2}\, \cos(s_+), ~ ~X_{\I R}={L\over 2}\, \cos(s_-) ~\to~ X_{\I }=L \cos({\tau\ov L})
\cos({\s\ov L}) 
\nonumber \\
Y_{\I L} &=& {L\over 2}\sin(s_+), ~ ~Y_{\I R}=- {L\over 2}\sin(s_-)~ \to~ Y_{\I }=L \sin({\tau\ov L})\cos({\s\ov L})
\label{sss}
\eea
\bea
X_{\II L}&=&{A\ov 2}+{L\over 2}\, \cos(s_++\alpha), ~  X_{\II R}={A\ov 2}+{L\over 2}\, \cos(s_--\alpha) 
\nn\\
~&\to &~ X_{\II }=A+L\cos({\tau\ov L}+\alpha)\cos({\s\ov L}) 
\nonumber \\
Y_{\II L}&=&{B\ov 2}-{L\over 2}\, \sin(s_++\alpha), ~ Y_{\II R}={B\ov 2}+{L\over 2}\,\sin(s_--\alpha) 
\nn\\
~&\to& ~  Y_{\II }=B-L\sin({\tau\ov L}+\alpha)\cos({\s\ov L}) 
\label{yyy}
\eea
$A,B$ and $\alpha $ are constants parametrizing the center of mass
coordinate
of the string II and its relative orientation. We take $A>0$.

The open strings are parametrized by $0\leq\s\leq \pi L$. Their energy 
is $E={1\ov 2\pi\alpha'}\int_0^{\pi L}d\s\p_\tau X^0=L/\alpha'$. 
We assume that the two strings interconnect at $\tau =0$.
They intersect at $\s_0$ and $\s'_0$ respectively.
The two strings $\I,\II$ of equal length recombine forming two strings
$a,b$ 
(or $c,d$) of different lengths forming  some kink. 

The intersection equations are:
\bea
0 = B- L \sin(\alpha)\cos({\s'_0})~&\to & ~ L\cos({\s'_0})={B\ov \sin(\alpha)} \\ \nonumber
L \cos(\s_0)= A+L \cos(\alpha)\cos(\s'_0)~&\to &~ L \cos(\s_0)=A+B{\cos(\ga)\ov \sin(\ga)}
\label{intersection}
\eea
A necessary condition for the strings to intersect at $\tau =0$ is $(A-L)^2+B^2\leq L^2$. 
The intersection equations
$\vec X_{\II L}(\s'_0)+\vec X_{\II R}(\s'_0)= \vec X_{\I L}(\s_0)+\vec X_{\I R}(\s_0)$
imply
\be
\vec X_{\II L}(\s'_0)-\vec X_{\I L}(\s_0)=\vec X_{\I R}(\s_0)-\vec X_{\II R}(\s'_0)\equiv \vec Q
\ee
Now we consider one of the two cases of interconnection shown in fig. 2.

The two open strings $\vec X_{a,b}(\s,\tau)$ after interconnection will be described by a world-sheet parameter $\s$ with interval of size $\Delta_a\s=\s_0+\pi-\s'_0$ and $\Delta_b\s=\s'_0+\pi-\s_0$ respectively
(the periodicity interval for the Left and Right part being the double of the above). Their energy 
is $L(\pi+\s_0-\s'_0)/2\pi\alpha'$ and $L(\pi+\s'_0-\s_0)/2\pi\alpha'$.

We will find that $\vec X_{a,b}(\s,\tau)$ have momentum and, as in Appendix A,
we will write
\be
\vec X_{a,b;L,R}(s)=[\vec X_{a,b,L;R}(s)\mp\vec k_{a,b}s]_{(0,2\Delta_{a,b}\s )}\pm\vec k_{a,b}s
~~~(+ ~ {\rm for} ~ L~ {\rm and} ~ - ~ {\rm for} ~R)
\label{ggg}
\ee
where we define the periodic function $[f(s+2\Delta\s)]_{(0,2\Delta\s )}=[f(s)]_{(0,2\Delta\s )}$.
In physical units the momenta are $\vec p_{a,b}=\vec k_{a,b}\Delta_{a,b}\s/\pi\alpha'$.

\vskip0.2cm

The string $a$ is   $~~~({\rm period}~2\Delta_a\s=2\pi-2\s'_0+2\s_0)$
\be
\vec X_{a L,R}(s) =
\begin{cases}
\vec X_{\I L,R} (s)\ , &  -\s_0\leq s\leq \s_0 \cr
\vec  X_{\II L,R} (s-\s_0+\s'_0)\mp \vec Q\ , &
~\s_0\leq s\leq 2\pi-2\s'_0+\s_0 \cr
\end{cases}
\label{ddd}
\ee
Further: 
\be
\vec k_a={1\ov 2\Delta_a\s}(\vec X_{\II L}(2\pi-\s'_0)-\vec Q-\vec X_{\I L}(-\s_0) )
=-{1\ov 2\Delta_a\s}(\vec X_{\II R}(2\pi-\s'_0)+\vec Q-\vec X_{\I R}(-\s_0) )
\nonumber
\ee
Explicitly
\be
k^x_a={L\ov 2\Delta_a\s}\sin(\ga )\sin(\s'_0 ) \ ,\qquad
k^y_a={L\ov 2\Delta_a\s}(\sin(\s_0)+\cos(\ga )\sin(\s'_0 ))\ ,
\nonumber
\ee
such that
$[\vec X_{a L,R}(s)\mp \vec k_a (s+\s_0)]$ take the same value at $s=-\s_0$ and $s=2\pi-2\s'_0+\s_0$. 

\vskip0.2cm

The open string $\vec X_a(\s,\tau)=\vec X_{a L}(\s+\tau)+\vec X_{a R}(\s-\tau)$ is defined for  
$0\leq\s\leq\Delta_a\s$. One can check that $\p_{\s}\vec X_a=0$ for $\s=0,\Delta_a\s$ and any $\tau$. 

The string $\vec X_b$ is obtained by interchanging $X_{\I L,R}\leftrightarrow X_{\II L,R}$ and
$\s_0\leftrightarrow \s'_0$ in the formulas for $\vec X_a$.
We get $\Delta_b\s=\pi+\s'_0-\s_0$,
\be
k^x_b=-{L\ov 2\Delta_b\s}\sin(\ga )\sin(\s'_0 ) \ ,\qquad
k^y_b=-{L\ov 2\Delta_b\s}(\sin(\s_0)+\cos(\ga )\sin(\s'_0 ))\ .\nonumber
\ee
As expected, the momenta are equal and opposite, that is
$\vec p_{a}=\vec k_{a}\Delta_{a}\s/\pi\alpha'=-\vec p_{b}=-\vec k_{b}\Delta_{b}\s/\pi\alpha'$
and the energy is conserved 
$$E=2{1\ov 2\pi\alpha'}\int_0^{\pi }d\s\p_\tau X^0=
{1\ov 2\pi\alpha'}\int_0^{\Delta_a\s }d\s\p_\tau X^0+
{1\ov 2\pi\alpha'}\int_0^{\Delta_b\s }d\s\p_\tau X^0\ .
$$

Note that the interconnecting equations contain as a particular case the joining
considered in the previous section, which is formally obtained for $\s_0=0 $
and $\s_0'=\pi $.

The other possible pair shown in fig. 2, $\vec X_{c,d}(\s,\tau)$, can be constructed in a similar way.
The strings will have energy 
$E_{c,d}={\Delta_{c,d} \s\over 2\pi\a '}L$, with $\Delta_c\s =\s_0+\s_0'\ ,\ \Delta_d\s =\pi - \s_0-\s_0' $.

\subsubsection{Black hole events}

Having the solutions of the two outgoing strings after the interconnection,
we now consider their evolution and study the possible black hole formation.

We first explore the possibility that the whole mass of the outgoing string 
collapses to a size less than the Schwarzschild radius. 
Specifically,
in this subsection we examine two conditions for black hole formation:
\smallskip

\noindent 1)  At some time during the evolution the average size of the
string,
\be
\bar R^2\equiv {1\over\Delta\s_{a,b}} \int_0^{\Delta\s_{a,b}}d\s 
R^2(\s )\ ,
\ee
\be
R^2(\s )= ( X_{a,b}(\s,\tau )- X_{a,b}^{\rm CM})^2+ 
( Y_{a,b}(\s,\tau )-  Y_{a,b}^{\rm CM})^2\ ,
\nn
\ee
is less than the Schwarzschild 
radius $R_s = 2G M$, where $M$ is the mass of one of the outgoing
strings $a$ or $b$.

\smallskip

\noindent 2) At some time during the evolution all points of the
string lie within the  Schwarzschild 
radius, i.e. $R(\s )<R_s \ $ for all $\s $.

\medskip

The masses of the strings $a$ and $b$ are given by
\be
M_{a,b}=L {\Delta_{a,b}\s \over 2\pi\a'} \sqrt{ \big( 1- 4\vec
k_{a,b}{}^2 \big) } \ .
\ee
It is convenient to express $R_s$ as
\be
R_s = 2 (G\mu ) {M\over \mu } \ ,\qquad \mu={1\over 2\pi\a '}\ .
\ee
The fundamental string has a tension $\mu $ whose value could be anywhere between the TeV scale and the Planck scale.
In brane inflation models, one expects a narrower range
$10^{-12}< G\mu< 10^{-6}$.

We have followed the evolution of the strings after interconnection in
$N_e$ events taking random values for the center of mass coordinates $A,B$ and for the relative orientation
$\alpha $ (within the intersection range).
We have seen that, when the strings shrink to a minimum size which is much smaller
than the initial size, they typically  have a shape
describing an incomplete circle.\footnote{In the case of a mass
distributed on a circle, a ``Laplace" Schwarzschild radius $R_s$
can be defined by requiring that the gravitational potential at the center is equal to $c^2/2$ ($c$ speed of light). This gives $R_s=2GM$, as in the case of the sphere studied by Laplace.}

The number of black hole events $N_{bh}$ depend on the value of $G\mu $.
Table 1 summarizes our results. We see that the condition $R(\s)<R_s$
for all $\sigma $ gives less black hole events. This is due to cases
where a small tail of the string lies outside the Schwarzschild radius.

\begin{table}[htbp]
\label{bhs}
\centering
\begin{tabular}{||l|l|l|l||}
$N_e$ & $G\mu$   & $N_{bh}$ ($\bar R<R_s$)   & $N_{bh}$ ($R(\s)<R_s$) \\
\hline
10000 & $ 10^{-2}$ & 1900 -- 2000 & 1100 -- 1200   \\ 
10000 & $ 10^{-3}$ & 300 -- 320  & 95 -- 110  \\ 
10000 & $ 10^{-4}$ & 40 -- 46  & 1 -- 3  \\ 
50000 & $ 10^{-5}$ & 20 -- 30  & 0 -- 4  \\ 
50000 & $ 10^{-6}$ & 3 -- 5   & 0 \\ 
\end{tabular}
\parbox{5in}{\caption{Number of black hole events in $N_e$ string collisions. }}
\end{table}



We also observe that the distribution of black hole
events in the region of possible parameters for the center of mass
coordinates $A$ and $B$ 
is nearly homogeneous. 
 {}From the data of table 1 one sees that $N_{bh}$ (computed with
 either criterium) has a power-like dependence with $G\mu $.


A typical black hole event is shown in Figure 3. 
The string after the interconnection has a kink at $\tau=0$, which
then separates into two kinks moving in opposite directions.
If the two pieces that form the string have
a comparable size, then the strings after the interconnection will
have a small angular momentum. This is typically the situation leading to 
a contraction of the strings to very small size. 

This figure, however, does not give information on how the mass is
distributed. In fact, as we will see in the next Section, at some
$\tau =\tau_0$ an important fraction of the mass is always
concentrated in one point. This fact, which is not possible to see in
fig. 3 (but can be seen in fig. 4), implies that all cases of this collision of Jmax + antiJmax strings should lead to black hole formation.
In particular, this indicates that the cross-section for the scattering of 
two long strings to form a black hole is essentially given by  the
geometric  area of the overlap of the two strings, times
$g_o^4=g_s^2$, where $g_s$ being the closed string 
coupling constant.


\bigskip

\begin{figure}[ht]
\centerline{
\epsfig{file=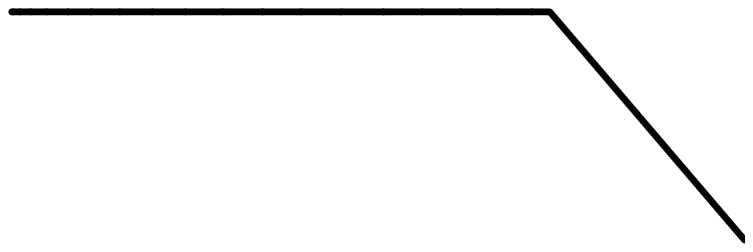,width=.4\textwidth}\ 
\epsfig{file=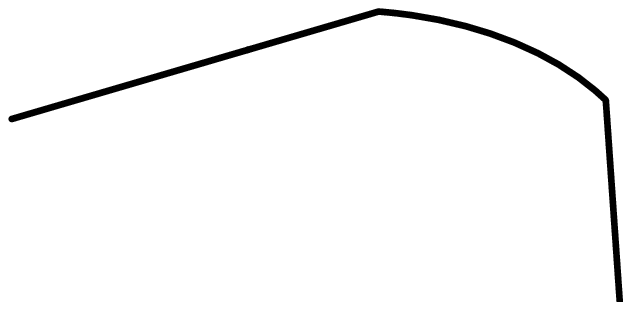,width=.4\textwidth}\  
\epsfig{file=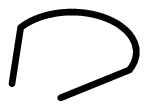,width=.4\textwidth}
 }
\centerline{
\epsfig{file=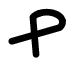,width=.4\textwidth}\  
\epsfig{file=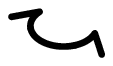,width=.4\textwidth}\  
\epsfig{file=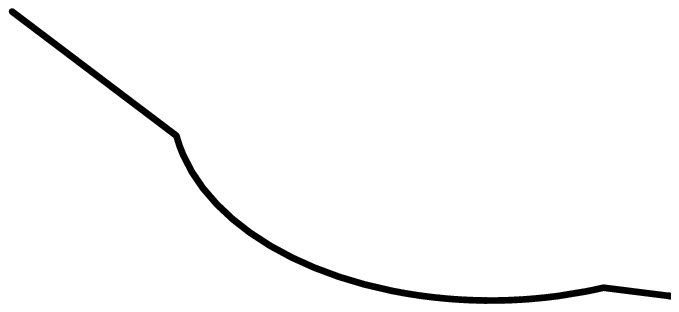,width=.4\textwidth}
 }
 \caption{\it A string solution resulting from interconnection possibly leading
to gravitational collapse, after shrinking by its own classical
 evolution. A generic feature 
is the formation of two kinks moving in opposite directions along the
 string. The figures are not schematic, they are obtained using
 the exact classical evolution of the string.
}
\end{figure}

\bigskip

\subsubsection{Inevitable Collapse in a generic  $J_\I =-J_\II $ case}

In section 3.1 we have already seen that a black hole will form in the
case of the {\it joinings} of Jmax and antiJmax strings.
In that case, the string that results from 
the joining process shrinks, becoming a point at
$\tau=\pi/2$. That is, just by the evolution dynamics in flat space, all
the mass concentrates in a region of 
zero size. We argued that energy loss by
radiation or breaking is negligible and inclusion of gravitational effects
will  reinforce the shrinking and finally a black hole will form.    

In the case of interconnection, 
we first observe that the same phenomenon of the complete shrinking of
the whole mass occurs when the interconnection takes place at $\s_0=\s_0'=\pi/2$ and $\alpha=0$
(with zero momentum in the transverse direction), that is, when the interconnection takes place 
in the middle point at zero angle. In this case, the interconnected strings have momentum along $Y$.  
Again, the string after interconnection shrinks to zero size 
at $\tau_0 =\pi/2$, and the same previous argument applies with
the conclusion that a black hole will form.       

The underlying mechanism is the cancellation of the dependence in $\sigma$ of the Left part with the Right part:
in more detail, for that value of $\tau=\tau_0$, the Left piece that, by construction, equals the Left piece of 
$X_{\II}$, cancels with the Right piece that equals $X_{\I}$ and 
{\it viceversa}.   

Let us now consider the slightly more general case in which 
$\s_0=\s_0'=\pi/2$, but $\alpha\neq 0$.
 The periodicity interval in $\sigma$ of the interconnected strings is again $\Delta\s =2\pi$.
We find that at $\tau_0 =\pi/2-\alpha/2$ the dependence in $\sigma$ cancels in the intervals
$-\pi+\alpha/2\leq\sigma\leq -\alpha/2$ and  $\alpha/2\leq\sigma\leq\pi -\alpha/2$.
Since the energy, and the mass, of the string is uniformly distributed in $\sigma$,
we see that for small $\alpha$  almost all of the mass shrinks to zero size. Also in this case
a black hole will form
with a mass of the same order of the mass of the incoming 
strings. In fact, the dimension of the incoming strings sets the overall scale, and thus the result
holds for an arbitrary value of the mass.\footnote{In particular, this means
that for strings with masses much larger than the Planck mass (as it
is obviously the case for astrophysical cosmic strings) quantum
gravity effects can be ignored, since the Schwarzschild radius will be much larger than the Planck
scale.}

By the same reasons, even when $\alpha$ is not small, a finite
fraction of the (arbitrarily large) mass
will shrink to zero size and form a black hole.

One can further investigate the general case of generic values of 
$\s_0,\, \s_0',\ \alpha$
(recall that the periodicity interval in $\sigma$ of the string $a$ is 
$2\Delta_a=2\pi\s -2\s_0'+2\s_0$ 
and of the string $b$ it is $2\Delta_b\s =2\pi-2\s_0+2\s_0'$). 
The resulting strings have in general linear momentum in the $XY$ plane 
and angular momentum as well. 
Taking for instance the string $b$, we find that a finite fraction of
its arbitrarily large mass
shrinks to zero size at 
$$
\tau_0=  \begin{cases}
{\pi \over 2}- {\alpha\over 2} + {1\over 2} (\s_0'-\s_0)  \ & {\rm if}\ \Delta_b\s > \alpha \cr
{\pi \over 2}- {\alpha\over 2} -{1\over 2} (\s_0'-\s_0)\     & {\rm if}\ \Delta_b\s    < \alpha \cr
\end{cases}
$$
This fraction is finite except for marginal values of $\s_0,\ \s_0',\ \alpha$,
and therefore an arbitrarily large black hole
is the generic result of the interconnection of arbitrarily large
strings of equal and opposite maximal angular momentum.

The above results can be numerically tested by constructing $\vec
X_{b,L}$ and $\vec X_{b,R}$ with the prescription given in
the previous section, and then  plotting in $\s$ both $X_b[\s,\tau_0]= X_{b,L}[\s+\tau_0]+X_{b,R}[\s-\tau_0]$
and $Y_b[\s,\tau_0]= Y_{b,L}[\s+\tau_0]+Y_{b,R}[\s-\tau_0]$. 
One can see that, in two intervals in $\s$, 
both $X$ and $Y$ are constant. A sample is shown in Fig. 4. 
One can check that for generic values of  $\s_0,\, \s_0',\ \alpha$  
the figures are similar.

\bigskip

\begin{figure}[ht]
\centerline{
\epsfig{file=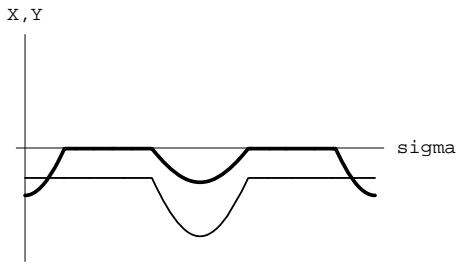,width=.4\textwidth}
 }
 \caption{\it For generic values of the center of mass parameters
parameters $A, B$ and relative orientation $\alpha $, part of the string
shrinks becoming
a point during the time evolution. In the figure, these are the values
of $\sigma $ for which both $X$ (thicker line) and $Y$ (thiner)
are constant.
}
\end{figure}

\bigskip

Finally, let us consider the case in which the interconnecting strings
have, in addition to angular momentum, equal and opposite linear momentum 
along the transverse $Z$ direction. 
In this case the interconnected strings will in general also stretch 
 in the $Z$ direction (periodically in $\tau$,  if one forgets gravity) and therefore 
the finite fraction of the string described above will not shrink exactly to zero size at
$\tau_0$. In order to conclude that a black hole 
will still form we have to compare the elongation in $Z$ with the 
Schwarzschild radius $R_s$.
The maximum elongation in $Z$ is of order $T\, v$, where $T\sim \ell $ is
the period of the motion, $\ell $ is the length of the strings and
$v$ is the relative velocity between the centers of mass.
Like before, the overall string scale factors out and we get the 
condition, for a non relativistic center-of mass motion of the string,
that the ratio of the relative velocity  $v$ between the strings to the
velocity of light ($c=1$) 
should be smaller than $G\mu$ times 
some number of order 1 depending on $\s_0,\ \s_0'$. 

As already mentioned, gravitational collapse of $J_\I=-J_\II$ strings
for generic initial data 
is expected to occur also in the 
case of the more stable open string which oscillates in the transverse
dimension (\ref{squash}), provided the string size $\ell_{\rm extra }$ 
in the extra dimensions is a small fraction of the overall size
$L$, 
but still much larger than $l_s=\sqrt{\a' }$ to ensure stability. More precisely,  
$\mu^{-1/2} \ll \ell_{\rm extra }< G\mu L $ or ${l_s\over L}\ll \theta
<G\mu $.

\subsubsection{Interconnection of rotating closed strings on $M^ 4\times S^ 1$}

Now consider the interconnection of the long-lived closed strings of Section 2.1.
Let us consider the case of two strings having the same winding in $W$.
For the $X,Y$ coordinates, the solutions of the strings I and II
are the same as in eqs. (\ref{sss}) and (\ref{yyy}), now with $\s \in [0,2\pi)$.
Similarly, the solutions after interconnection are the same as in eqs. (\ref{ggg}) and (\ref{ddd}).
The closed strings $\vec X_{a,b}(\s,\tau)=\vec X_{a,b L}(\s+\tau)+\vec X_{a,b R}(\s-\tau)$ are defined for  
$0\leq\s\leq 2\Delta_{a,b}\s $.

For these closed strings, the interconnection takes place at two points $\s_0$ and $2\pi-\s_0$ for the string I
and  $\s_0'$ and $2\pi-\s_0'$ for the string II. 

The intersection in the $W$ coordinate
requires that $W_\I(\s_0)=W_{\II }(\s_0')$ and  $W_\I(2\pi-\s_0)=W_{\II }(2\pi-\s_0')$,
where we take $W_\I=R\s_\I \ ,\  W_{\II}=R\s_\II $.
This implies $\s_0=\s_0'$ and, in turn, $2\Delta_{a,b}\s =2\pi $, consistently with 
the fact that the strings $a$ and $b$ have both winding numbers equal to 1.

The results of the previous Section tell us that for 
$\tau=\tau_0=\pi/2-\alpha/2$ a finite fraction of the arbitrarily large mass 
of the string undergoes gravitational collapse, for generic values of $\alpha$.  

Classically, the condition $\s_0=\s_0'$ implies that rather than a geometric area we get a one-dimensional 
``cross-section". Quantum mechanically, however,
the interconnection process can also take place if 
the interconnecting points are separated by a distance of order 
$\sqrt{\a '}$.
Therefore the cross section for the collapse of a 
finite fraction of the string  
will be of the order of the string length times $\sqrt{\a '}$, times $g_s^4$.


\section*{Acknowledgements}

We would like to thank D.~Amati and J.~Garriga for  useful discussions.
This work is
supported in part by the European
EC-RTN network MRTN-CT-2004-005104. J.R. also acknowledges support by MCYT FPA
2004-04582-C02-01 and CIRIT GC 2005SGR-00564.

\setcounter{section}{0}
\setcounter{subsection}{0}

\setcounter{equation}{0}

\appendix{String splitting and joining: \\ general formalism}
\setcounter{equation}{0}

\subsection{Splitting of Closed and Open Strings}

Consider first the splitting of 
a closed string. The initial closed string is
 described by the solution
\be
X_0=\a' M \tau\ ,\qquad \vec X_R=\vec X_R(\s_-)\ ,\qquad  
\vec X_L=\vec X_L(\s_+)\ ,
\ee
with $\s_\pm=\s\pm\tau $ and $\s\in [0,2\pi)$. 
In this gauge $X_0=\a' M \tau $, the  
Virasoro constraints become:
\be
\big( \p_s \vec X_{R,L}(s) \big)^ 2={1\over 4} (\a' M)^ 2\ .
\ee
The momentum of the string is $\a' \vec p=\vec p_R+\vec p_L$, with
\be
\vec p_R= -{\vec X_R(2\pi )- \vec X_R(0 )\ov 2\pi }\ ,\qquad
\vec p_L= {\vec X_R(2\pi )- \vec X_R(0 )\ov 2\pi }\ .
\ee
For a string with winding $w$ around a compact dimension $X$ of radius
$R$, 
one has  $ p_R={1\ov 2}(\a' p-
 wR)$,
$ p_L={1\ov 2}(\a' p+  wR)$,
so that $p_L-  p_R= wR$.
If $ p_{R,L}\neq 0$ then  $X_{R,L}(s)$ is not periodic.
In such a case, one can define a periodic function by substracting
the non-periodic part,  
\bea
X_R(s) &=& \Big(  X_R(s) +  p_R s\Big)-   p_R s\ ,
\nn\\
X_L(s) &=& \Big(  X_L(s) - p_L s\Big)+   p_L s\ .
\eea
The functions within the parentheses $ \Big( \cdots \Big)$ are continuous
 --~but kinks are allowed~-- 
and periodic by definition . 
This simple observation will be useful for the construction
below.

\medskip

Now we assume that at $\tau =0$ there is a contact between two points
of the closed string and the string breaks into two fragments
$\vec X_\I,\ \vec X_\II $,
\be
\vec X_\I= \vec X_{\I R}(\s_-) +\vec X_{\I L}(\s_+)\ ,\qquad 
\vec X_\II= \vec X_{\II R}(\s_-) +\vec X_{\II L}(\s_+)\ .\
\ee
The fragment solutions are uniquely
determined
by the condition that the functions $X$ and their first time derivatives
are continuous at $\tau =0$.
The first fragment is defined to be the piece of the string with
$\s_1 <\s <\s_2$ while the second fragment is the remaining
piece $\s_2<\s<2\pi+\s_1$.
The outgoing strings will carry in general non-zero momentum. Since it
is conserved, this can be computed at $\tau =0$.
They are given by $\a'\vec p_\I=\vec p_{\I R}+\vec p_{\I L}$,
\bea
\vec p_{\I R} &=& -{\vec X_{R}(\s_2 )- \vec X_{R}(\s_1 )\ov 2\pi }\ ,\qquad
\vec p_{\I L}= {\vec X_{L}(\s_2 )- \vec X_{L}(\s_1 )\ov 2\pi }\ ,
\nn\\
\vec p_{\II R} &=& -{\vec X_{R}(2\pi+\s_1 )- \vec X_{R}(\s_2 )\ov 2\pi }\ ,\qquad
\vec p_{\II L}= {\vec X_{L}(2\pi+\s_1 )- \vec X_{L}(\s_2 )\ov 2\pi }\ .
\label{momenta}
\eea

Consider the general case where there may be compact dimensions
of radii $R_i$, $i=1,...,D-1$. This includes the uncompact $R=\infty $ case. 
The breaking is possible if $ X_i (\s_1,0)=  X_i (\s_2,0)$ mod
$n_iR_i$ where $n_i$ are integers
(there is no summation over $i$). For uncompact dimensions, $n_i=0$.
This condition can be written as
\be
 X_{i\, L} (\s_1,0)- X_{i\, L} (\s_2,0)=-  X_{i\, R} (\s_1,0)+
 X_{i\, R}
(\s_2,0) + 2\pi n_i R_i
\label{condition}
\ee
or
\be
p^\I _{iL}= p^\I _{i R}+ w_i^\I  R_i
\ee
with $w^\I_i= n_i$. The breaking occurs if this condition can be satisfied 
for all $i$, for some $\s_2$ and $\s_1$ 
(in some cases, it could be that there are several
solutions, i.e. many contact points).
Similarly, since   $p_{iL}-  p _{i R}= w_i  R_i $\ ,
\be
 p^\II _{iL}= p^\II _{i R}+ w_i^\II  R_i\  ,\qquad
 w_{i}^\I + w_{i}^\II= w_i\ .
\ee

The energies are
\be
E_\I ={(\s_2-\s_1)\over 2\pi} M\ ,\ \ \ \ E_\II=M-E_\I=
{(2\pi+\s_1-\s_2)\over 2\pi} M\ .
\ee
The masses of each of the outgoing fragments are then given by
\bea
M_\I^ 2=M^ 2 {(\s_2-\s_1)^ 2\over 4\pi^ 2}- {\vec p_\I}{} ^ 2\ ,
\nn\\
M_\II^ 2=M^ 2 {(2\pi+\s_1-\s_2)^ 2\over 4\pi^ 2}- {\vec p_\II}{}^ 2\ .
\label{masses}
\eea
This defines $M_\I $ in terms of $M_\II $ and in
terms of 
the quantum numbers of the original string.

The initial condition 
uniquely determines the outgoing solutions to be given by
\bea
\vec X_{\I R}(s) &=& \big[ \vec X_R(s) - {\vec  X_R(s_2)-\vec
    X_R(s_1)\over s_2-s_1 } \ s \big]_{(s_1,s_2)} + {\vec  X_R(s_2)-\vec
    X_R(s_1)\over s_2-s_1 } \ s  \ ,
\nn\\
\vec X_{\I L}(s) &=& \big[ \vec X_L(s) - {\vec  X_L(s_2)-\vec
    X_L(s_1)\over s_2-s_1 } \ s \big]_{(s_1,s_2)} + {\vec  X_L(s_2)-\vec
    X_L(s_1)\over s_2-s_1 } \ s  \ ,
\eea
where we have introduced the symbol $\big[f(x)\big]_{(a,b)}\equiv\hat
f(x)$ as the periodic function  defined by $\hat f(x+n(b-a))= \hat f(x)\ ,\
x\in [a,b)$ and $n$ is an integer.

Similarly, the second fragment is described by the solution 
\bea
\vec X_{\II R}(s) &=& \big[ \vec X_R(s) - {\vec  X_R(2\pi+s_1)-\vec
    X_R(s_2)\over 2\pi+s_1-s_2 } \ s \big]_{(s_2,2\pi +s_1)} + {\vec  X_R(2\pi+s_1)-\vec
    X_R(s_2)\over  2\pi+s_1-s_2 } \ s  
\nn\\
\vec X_{\II L}(s) &=& \big[ \vec X_L(s) - {\vec  X_L(2\pi+s_1)-\vec
    X_L(s_2)\over  2\pi+s_1-s_2 } \ s \big]_{(s_2,2\pi +s_1)} + {\vec  X_L(2\pi+s_1)-\vec
    X_L(s_2)\over  2\pi+s_1-s_2 } \ s  \nn
\eea
By the above equations we have required the Left and Right sectors of the string to be the same at $\tau=0$ as functions of the world-sheet parameter $\s$ in the interval $0\leq \s\leq 2\pi$. This implies also the continuity of the first derivative in $\tau$ at $\tau =0$ since 
$\p_{\tau}\vec X_{\tau =0}=\p_{\s}\vec X_L(\s )-\p_{\s}\vec X_R(\s )$.

It is convenient to rescale the $s$ variable to have $2\pi $ periodic
functions.
We define, for the fragment I, 
\be
s = \hat s \ {(s_2-s_1)\over 2\pi }+ s_1\ ,
\ee
whereas for the fragment II
 \be
s  = \hat s \ {(2\pi+s_1-s_2)\over 2\pi }+ s_2\ .
\ee
Note that this implies that both $\sigma$ and $\tau$ get rescaled, and that 
we imposed continuity of the derivative with respect to the unrescaled $\tau$.

The solutions are then as follows:
\bea
\vec X_{\I R}(s) &=& \big[ \vec X_R(\hat s \ {(s_2-s_1)\over 2\pi }+
s_1 ) + \vec p_{\I R}\hat s  \big]_{(0,2\pi)}-\vec p_{\I R}\hat s
\nn\\
\vec X_{\I L}(s) &=& \big[ \vec X_L(\hat s \ {(s_2-s_1)\over 2\pi }+  s_1 ) -\vec p_{\I L}\hat s  \big]_{(0,2\pi)}+\vec p_{\I L}\hat s
 \nn
\eea
\bea
\vec X_{\II R}(s) &=& \big[ \vec X_R(\hat s \ {(2\pi+s_1-s_2)\over 2\pi }+  s_2 ) +\vec p_{\II R}\hat s  \big]_{(0,2\pi)}-\vec p_{\II R}\hat s
\nn\\
\vec X_{\II L}(s) &=& \big[ \vec X_L(\hat s \ {(2\pi+s_1-s_2)\over 2\pi }+  s_1 ) -\vec p_{\II L}\hat s  \big]_{(0,2\pi)}+\vec p_{\II L}\hat s
 \nn
\eea
The above construction holds also for open strings (see \cite{IR2}). 
In this case one has simply to remember that for an open string
$ X_R^\mu (\s_{-})=X_L^\mu (-\s_{-})$ and that the interval in $\s$ is $[0,\pi)$. 

It is useful to  explicitly separate the momentum term as follows: 
\be
X_L^\mu (\s_{+})= F^\mu(\s_{+})+\a' p^\mu \s_+ +c^\mu, ~ ~
X_R^\mu (\s_{-})= F^\mu(-\s_{-})-\a' p^\mu \s_- -c^\mu
\ee 
where $c^\mu$ is a constant and $F^\mu$ is periodic i.e. $F^\mu(s)=F^\mu(s+2\pi)$  
so that 
\be
X^\mu_{open} (\s,\tau )=F^\mu (\s+\tau )+F^\mu (-\s+\tau ) +2\a'  p^\mu \tau~. \label{open}
\ee

\subsection{Joining of Open Strings.}

Now consider two {\it open} strings I and II  in the CM frame with energies $E_{\I}$ and $E_{\II}$.
We assume that at $\tau=0$ one end of the string I gets in contact
with one end of the string II, that is $\vec X_{\I}(0,0)=\vec X_{\II}(\pi,0) $. 
If the two strings join making one final string $\vec X$, that one will have a mass $M=E_{\I}+E_{\II}$.

Therefore since $(\p_{s_\I} \vec X_{\I L,R})^2=(\a' E_{\I})^2, ~~ (\p_{s_\II} \vec X_{\II L,R})^2=(\a' E_{\II})^2$
and for the final string $(\p_{ s}\vec  X_{ L,R})^2=(\a'  M)^2$, the wordsheet parameter 
$0\leq s\leq 2\pi$ of the final string must be related to the ones of the joining strings 
by $s_{\I,\II} ={M\ov E_{\I,\II}} s+c_{\I,\II}$.

As it has been said in the splitting case, the matching requirement is equivalent to requiring that the resulting 
$\vec X_{L,R}(s)$ is piecewise identical to $\vec X_{\I L,R}(s_{\I})$ and $\vec X_{\II L,R}(s_{\II})$. 
Therefore we get $\vec X$ by the following construction:
\be
\vec X_{L,R}(s) = 
\begin{cases}
\vec X_{\II L,R} ({M\ov E_{\II}}s)\  & -{E_{\II}\ov M}\pi\leq s< {E_{\II}\ov M}\pi \cr
\vec X_{\I L,R} ({M\ov E_{\I}}(s-{E_{\II}\ov M}\pi))\  & \ {E_{\II}\ov M}\pi\leq  s< 2\pi-{E_{\II}\ov M}\pi\cr
\end{cases}
\label{kkp}
\ee 
Since $\vec X_{L,R}(s) $ must be $2\pi$ periodic, joining is only possible if \\
$$
\vec X_{\II L,R} (-\pi)=\vec X_{\I L,R} (({M\ov E_{\I}}-{E_{\II}\ov E_{\I}})2\pi)\ .
$$

 As a particular case, consider two strings of equal mass, carrying equal and opposite 
momenta, described by the solutions
\bea
&X_{0\I} =2\a' E \tau \ , & ~
\vec X_{\I L}(s) = \vec F_{\I }(s) +\a'\vec p (s- {\pi\over 2}) ~,\  ~
\vec X_{\I R}(s) = \vec F_{\I }(-s)-\a'\vec p (s- {\pi\over 2}) \nonumber \\ 
&X_{0\II}=2\a' E \tau& 
\vec X_{\II L}(s) = \vec F_{\II }(s)  -\a'\vec p (s- {\pi\over 2}) ~,\
~ 
\vec X_{\II R}(s) = \vec F_{\II }(-s)+\a'\vec p (s- {\pi\over 2}) \nonumber \\ 
\eea
where $\vec F_{\I ,\II}(s)$ is $2\pi $-periodic  and  by assumption 
$(\p_s \vec F_{\I}+\a' \vec p)^2=(\p_s \vec F_{\II}-\a' \vec p)^2=(\a' E )^2$.
The joining condition 
$\vec X_{\I}(0,0)=\vec X_{\II}(\pi,0) $ implies $\vec F_{\I }(0)=\vec F_{\II }(\pi)$.

Explicitly, in this case $E_{\I}=E_{\II}=M/2$,
the resulting solution after the joining is given by
\be
\vec X_{L,R}(s) = 
\begin{cases}
\vec X_{\II L,R} (2s)\  & \ -{\pi\ov 2}\leq s< {\pi\over 2} \cr
\vec X_{\I L,R} (2s-\pi)\  & \ {\pi\over 2}\leq s< {3\pi\over 2}\cr
\end{cases}
\label{uuu}
\ee 
Note that $\vec X_{L,R}(-{\pi\over 2})=\vec X_{L,R}({3\pi\over 2})$ 
and that $X^\mu (\s,\tau )$ has the open string structure
(\ref{open}) with zero momentum.
Outside the interval $-{\pi\ov 2}\leq s<{3\pi\over 2}$, $~\vec X_{L,R}(s)$ is defined by
its periodic extension, i.e. by replacing $s$ by $\hat s= s-[s/2\pi]$. 

The resulting string  being periodic, after one period $\Delta\tau
=2\pi$ it comes back to the original configuration.
Being an open string,
it could split again at anytime. For example, at  $\Delta\tau
=2\pi$ it could split into the two original pieces $\vec X_{\I, \II}$ or else continue in its periodic motion.


\medskip

As an application,  consider now the case where 
two open strings with maximum angular momentum move in the
same clockwise sense. The solutions are
$X (\tau,\s )=X_{ L}(\tau+\sigma )+ X_{L}(\tau-\sigma )$
with
\be
X_{\I L}(s)= {L \over 2}\, \cos s ,\qquad  Y_{\I L}={L\over 2}\, \sin s\ ,
\ee
\be
X_{\II L}(s)= L+ {L \over 2}\, \cos s ,\qquad  Y_{\I L}={L \over 2}\, \sin s\ .
\ee
The main difference with respect to case of opposite angular momenta discussed in Section 3.1 is that,
at the moment of the joining at $\tau=0$, the ends 
are now moving with opposite velocities. In the previous case,
they were moving with the same velocity and, as a result, the string
which
resulted after the joining was smooth.
Now, because the attached ends are moving at opposite velocities,
a kink will be formed. This case illustrates that the formation of
kinks
in string joining (as in string splitting \cite{IR2}) is  generic,
since the generic situation is that the velocities of the two joined
ends are different (as vectors, the endpoints of open strings always move at the speed of light).

The solution after the joining can be constructed using eq. (\ref{uuu}).
Beside the formation of the kink, another
 interesting feature is that the open strings become folded
 cyclically during the evolution.

\subsection{Joining of  closed strings }

Consider two closed strings I and II described by the solutions
\bea
X_{0\I} &=& \a' E_\I \tau\ ,\qquad \vec X_{\I R}=\vec X_{\I R}(\s_-)\ ,\qquad  
\vec X_{\I L}=\vec X_{\I L}(\s_+)\ ,  
\nn\\
X_{0\II} &=& \a' E_\II \tau\ ,\ \ \ \vec X_{\II R}=\vec X_{\II R}(\s_-)\ ,\ \ \   
\vec X_{\II L}=\vec X_{\II L}(\s_+)\ .
\eea
We assume that at $\tau=0$ one point of the closed string I (say $\s_\I=0$) gets in contact
with one point of the closed string II (say $\s_\II=\pi $) , and the strings join.
The resulting solution is again uniquely determined by the assumption of
continuity of $X$ and $\dot X$ at $\tau=0$.
 
Now the solution after the joining is 
\be
X^\mu (\s,\tau )=X_L^\mu (\s+\tau )+X_R^\mu (-\s+\tau )\ ,
\ee
where $ X_L^\mu (\s+\tau )$ and $X_R^\mu (-\s+\tau )$ are determined by 
continuity of $X$ and $\dot X$ at $\tau=0$. When the energies are the same,
we find that $ X_L^\mu (\s+\tau )$ and $X_R^\mu (-\s+\tau )$ are given
by eq. (\ref{uuu}). When they are different, the solution is
constructed in a similar way as (\ref{kkp}).


\end{document}